\documentclass[10pt,twocolumn,english,prb,aps,superscriptaddress,longbibliography]{revtex4-2}
\usepackage[utf8]{inputenc}
\usepackage{amsmath}
\usepackage{amssymb}
\usepackage{graphicx}
\usepackage{xspace}
\usepackage{siunitx}
\usepackage{mathtools}
\usepackage{xcolor}
\usepackage{bbold}
\makeatletter
\@ifundefined{textcolor}{}
{\definecolor{BLACK}{gray}{0}
 \definecolor{WHITE}{gray}{1}
 \definecolor{RED}{rgb}{1,0,0}
 \definecolor{GREEN}{rgb}{0,1,0}
 \definecolor{BLUE}{rgb}{0,0,1}
 \definecolor{CYAN}{cmyk}{1,0,0,0}
 \definecolor{MAGENTA}{cmyk}{0,1,0,0}
 \definecolor{YELLOW}{cmyk}{0,0,1,0}
}

\usepackage[normalem]{ulem}

\usepackage{soul}
\usepackage{braket}
\usepackage[backref=none,
bookmarksnumbered=true,
bookmarks=true,
bookmarksopen=true,
colorlinks=true,
citecolor=blue,
linkcolor=blue,
anchorcolor=green,
urlcolor=blue,unicode=false]{hyperref}

\makeatother

\usepackage[english]{babel}

\newcommand{\comment}[1]{}

\newcommand{\Tr}{{\rm Tr\,}}
 
\def\cond{c} 
\def\val{v} 
 
\def\condop{c_{\mathbf k,\cond}} 
\def\valop{c_{\mathbf k,\val}} 
\def\condopp{c_{\mathbf k',\cond}} 
\def\valopp{c_{\mathbf k',\val}} 
\newcommand{\ehvacuum}{\ket{\text{IS}}} 
\newcommand{\vacuum}{\ket{\mathrm{vac}}} 
\newcommand{\GS}{\ket{\mathrm{GS}}} 
\newcommand{\rydberg}{\mathrm{Ry}^*} 
\newcommand{\bohrradius}{a_B^*} 

\newcommand{\nx}{n_{\text{ex}}} 
\newcommand{\nxatomic}{\tilde{n}_{\text{ex}}} 
\newcommand{\mux}{\mu_{\text{ex}}} 
\newcommand{\Nx}{N_{\text{ex}}} 

\newcommand{\auc}{A_{\text{UC}}}
\newcommand{\nsites}{N_{\text{UC}}}

\newcommand{\didv}{\ensuremath{\frac{{\mathrm d}I}{{\mathrm d}V}}\xspace}

\def\wfvector{\Phi}

\usepackage{ifthen}

\def\indexsolution{\alpha}

\newcommand{\deltazero}[1][]{
   \ifthenelse{ \equal{#1}{} }
    {\delta (E_{\mathbf k, \indexsolution}- eV)}
    {\delta (E_{\mathbf k, #1}- eV)}
}

\newcommand{\meanenergy}{\frac{\bar\epsilon^\cond_{\mathbf{k}}+\bar\epsilon^\val_{\mathbf{k}}}{2}}
\newcommand{\bcsenergy}{\sqrt{\left(\frac{\bar\epsilon^\cond_{\mathbf{k}}-\bar\epsilon^\val_{\mathbf{k}}}{2}\right)^2+\Delta_{\mathbf{k}}^2}}

\newcommand{\rtip}{\mathbf{r}_{\text{tip}}}

\newcommand{\wte}{{WTe\textsubscript{2}}}

\newcommand{\tise}{{TiSe\textsubscript{2}}}
\newcommand{\zrte}{{ZrTe\textsubscript{2}}}
\newcommand{\tanise}{Ta\textsubscript{2}NiSe\textsubscript{5}}
\newcommand{\tapdte}{Ta\textsubscript{2}Pd\textsubscript{3}Te\textsubscript{5}}

\newcommand{\bexpanded}[1][]{
   \ifthenelse{ \equal{#1}{} }
   {b_{\v{k}+\v{w}_{\lambda},\sigma,\lambda}}
    {b_{\v{#1}+\v{w}_{\lambda},\sigma,\lambda}}
}

\newcommand{\bexpandeds}[1][]{
   \ifthenelse{ \equal{#1}{} }
   {b_{\v{k}+\v{w}_{\lambda},s,\lambda}}
    {b_{\v{#1}+\v{w}_{\lambda},\sigma,\lambda}}
}
\newcommand{\aexpanded}[1][]{
   \ifthenelse{ \equal{#1}{} }
   {a_{\v{k},\sigma}}
    {a_{\v{#1},\sigma}}
}

\newcommand{\uka}[1][]{
   \ifthenelse{ \equal{#1}{} }
    {u_{\v{k},a,s}}
    {u_{\v{#1},a,s'}}
}
\newcommand{\ukasigma}[1][]{
   \ifthenelse{ \equal{#1}{} }
    {u_{\v{k},a,\sigma}}
    {u_{\v{#1},a,\sigma'}}
}
\newcommand{\phikasigma}[1][]{
   \ifthenelse{ \equal{#1}{} }
    {\phi_{\v{k},a,\sigma}}
    {\phi_{\v{#1},a,\sigma'}}
}
\newcommand{\phika}[1][]{
   \ifthenelse{ \equal{#1}{} }
    {\phi_{\v{k},a,s}}
    {\phi_{\v{#1},a,s'}}
}
\newcommand{\phikb}[1][]{
   \ifthenelse{ \equal{#1}{} }
    {\phi_{\v{k+w}_\lambda,b,s}}
    {\phi_{\v{#1+w}_\lambda,b,s'}}
}
\newcommand{\psikasigma}[1][]{
   \ifthenelse{ \equal{#1}{} }
    {\phi_{\v{k},a,\sigma}}
    {\phi_{\v{#1},a,\sigma'}}
}
\newcommand{\psika}[1][]{
   \ifthenelse{ \equal{#1}{} }
    {\phi_{\v{k},a,s}}
    {\phi_{\v{#1},a,s'}}
}
\newcommand{\psikb}[1][]{
   \ifthenelse{ \equal{#1}{} }
    {\phi_{\v{k+w}_\lambda,b,s}}
    {\phi_{\v{#1+w}_\lambda,b,s'}}
}

\newcommand{\ukb}[1][]{
   \ifthenelse{ \equal{#1}{} }
    {u_{\v{k+w}_\lambda,b,s}}
    {u_{\v{#1+w}_\lambda,b,s'}}
}

\newcommand{\uk}{u_{\mathbf{k}}}
\newcommand{\vk}{v_{\mathbf{k}}}

\begin{document}
\title{Tunneling signatures of interband coherence in dilute exciton condensates}

\author{Kry\v{s}tof Kol\'{a}\v{r}}
\affiliation{Department of Applied Physics, Aalto University School of Science, FI-00076 Aalto, Finland}
\affiliation{\mbox{Dahlem Center for Complex Quantum Systems and Fachbereich Physik, Freie Universit\"at Berlin, 14195 Berlin, Germany}}

\author{Felix von Oppen}
\affiliation{\mbox{Dahlem Center for Complex Quantum Systems, Fachbereich Physik, and Halle-Berlin-Regensburg}\\ Cluster of Excellence CCE, Freie Universit\"at Berlin, 14195 Berlin, Germany}

\date{\today}
\begin{abstract}
We theoretically investigate signatures of exciton condensation and the underlying interband coherence in scanning tunneling microscopy. We consider both monolayer and bilayer condensates in the regime of a dilute condensate of tightly bound excitons. For monolayer condensates, interband coherence is directly encoded in spatially oscillating contributions to the tunneling conductance, which break the underlying lattice symmetry. We show how
scanning tunneling microscopy allows one to extract the exciton wavefunction. For bilayer condensates, we show that the formation of the exciton insulator is signaled by the emergence of a characteristic peak in the tunneling conductance, which can be used to extract the (local) exciton density. Our results are based on analytical considerations using a systematic solution of the mean-field equations in powers of the exciton density as well as numerical calculations.
\end{abstract}
\maketitle

\section{Introduction}

Excitons  -- bound pairs of electron and hole --  are expected to form spontaneously  in the ground state of an intrinsic narrow-gap semiconductor if the binding energy $E_b$ of an isolated exciton exceeds the band gap $E_G$ between the conduction and valence bands
\cite{kohnJeromeExcitonicInsulator1967,riceHalperinExcitonicStateSemiconductorSemimetal1968,kozlovKeldyshCollectivePropertiesExcitons1968,maksimovKozlovMetalDielectricDivalentCrystal1965}.
When $E_G$ is close to $E_b$, the exciton density $\nx$ is small and the internal structure of the excitons becomes immaterial. They realize a gas of effectively  pointlike bosons, with the residual overlap of excitons inducing a finite exciton compressibility. This gas is expected to undergo Bose-Einstein condensation (BEC) at low temperatures \cite{kozlovKeldyshCollectivePropertiesExcitons1968}. The formation of the condensate affects the noninteracting band structure only mildly by increasing the gap  $E_g$. However, the associated  emergence of  interband coherence spontaneously breaks the separate charge conservation symmetries of the conduction and valence bands,
permitting exciton superflow under suitable conditions. As the exciton density increases, the system eventually crosses over into a Bardeen-Cooper-Schrieffer (BCS) regime, in which overlapping bare valence and conduction bands  spontaneously develop an excitation gap.

While exciton condensates have been studied for almost sixty years, recent years have seen a surge of interest in exciton insulators driven by the emergence of materials breakthroughs. Recent work on monolayers of \zrte{} \cite{chenGaoEvidenceHightemperatureExciton2023} and \wte{} \cite{wuJiaEvidenceMonolayerExcitonic2022,cobdenSunEvidenceEquilibriumExciton2022}, bilayer transition metal dichalcogenides \cite{shanMaStronglyCorrelatedExcitonic2021,makWangEvidenceHightemperatureExciton2019,wangQiThermodynamicBehaviorCorrelated2023}, and bulk \tapdte{} \cite{hasanHossainDiscoveryTopologicalExciton2023,qianHuangEvidenceExcitonicInsulator2024} presented evidence for exciton-insulator formation, improving upon earlier candidate materials such as \tanise{} \cite{ohtaSekiExcitonicBoseEinsteinCondensation2014,takagiHeTunnelingtipinducedCollapseCharge2021,takagiLuZerogapSemiconductorExcitonic2017,kimKimDirectObservationExcitonic2021} and \tise{} \cite{forroCercellierEvidenceExcitonicInsulator2007,abbamonteKogarSignaturesExcitonCondensation2017}. Unequivocal detection of an exciton insulator is challenging since it is a state with spontaneous mixing of conduction and valence bands \cite{rossnagelRossnagelOriginChargedensityWaves2011,nozieresComteExcitonBoseCondensation1982,riceHalperinExcitonicStateSemiconductorSemimetal1968}. If this band mixing occurs between bands with a finite momentum offset, it generically couples to the lattice \cite{rossnagelRossnagelOriginChargedensityWaves2011}, which can also drive band mixing. Recently, Refs.~\cite{rubioWindgatterCommonMicroscopicOrigin2021,gedikBaldiniSpontaneousSymmetryBreaking2023,chenGaoEvidenceHightemperatureExciton2023} doubt a purely excitonic origin of the states observed in \tanise{} and \tise{}, suggesting instead a lattice driven mechanism. Bilayer condensates in which conduction and valence bands are localized in separate layers, do not couple to the lattice \cite{zhuLittlewoodPossibilitiesExcitonCondensation1996,yakobsonGuptaHeterobilayers2DMaterials2020}, while the recent experiments on \zrte{} \cite{chenGaoEvidenceHightemperatureExciton2023} and \tapdte{} \cite{qianHuangEvidenceExcitonicInsulator2024}  rely on careful modelling to preclude a lattice-driven phase. Experiments on monolayer \wte{} \cite{wuJiaEvidenceMonolayerExcitonic2022,cobdenSunEvidenceEquilibriumExciton2022} argued based on transport anomalies and did not observe any lattice distortion, possibly as a result of the specific spin ordering \cite{parameswaranKwanTheoryCompetingExcitonic2021}.   

Diverse experimental techniques were used to probe candidate materials for exciton-insulator formation. Angle-resolved photoemission spectroscopy (ARPES) \cite{ohtaSekiExcitonicBoseEinsteinCondensation2014,yeomLeeStrongInterbandInteraction2019,gedikBaldiniSpontaneousSymmetryBreaking2023,forroCercellierEvidenceExcitonicInsulator2007,qianHuangEvidenceExcitonicInsulator2024,chenGaoEvidenceHightemperatureExciton2023} mapped band structures, and the spontaneous opening of a gap at small temperatures was taken as evidence. Bilayer systems, with extraordinary tunability afforded by bias and gate voltages \cite{macdonaldXieElectricalReservoirsBilayer2018,macdonaldZengElectricallyControlledTwodimensional2020}, allowed for thermodynamic  measurements of the exciton density as a function of the band gap $E_G$ \cite{shanMaStronglyCorrelatedExcitonic2021,makWangEvidenceHightemperatureExciton2019,wangQiThermodynamicBehaviorCorrelated2023}. Furthermore, bilayer systems allow for probing of exciton superflow, manifesting in perfect Coulomb drag \cite{wangQiPerfectCoulombDrag2023,makNguyenPerfectCoulombDrag2023} similar to quantum Hall bilayer condensates \cite{eisensteinEisensteinExcitonCondensationBilayer2014,westNandiExcitonCondensationPerfect2012,westKelloggVanishingHallResistance2004,huseTutucCounterflowMeasurementsStrongly2004}. Notably, global probes such as ARPES or thermodynamic and transport measurements are subject to averaging over sample inhomogeneities. 

A natural local probe of exciton insulators with high spatial and energy resolution is scanning tunneling microscopy (STM). STM has so far been used in the study of excitonic states in a limited way, primarily to confirm a gapped state \cite{wuJiaEvidenceMonolayerExcitonic2022,chenGaoEvidenceHightemperatureExciton2023,qianHuangEvidenceExcitonicInsulator2024,yeomLeeStrongInterbandInteraction2019,hasanHossainDiscoveryTopologicalExciton2023,takagiHeTunnelingtipinducedCollapseCharge2021}.
The usefulness of STM beyond establishing gapped states was highlighted by recent experiments. Experiments on flat-band graphene-based systems \cite{yazdaniLiuVisualizingBrokenSymmetry2021,nadj-pergeKimImagingIntervalleyCoherent2023,yazdaniNuckollsQuantumTexturesManybody2023} use related signatures to probe inter-valley coherence \cite{bernevigCalugaruSpectroscopyTwistedBilayer2021,zaletelHongDetectingSymmetryBreaking2021,bultinckKwanKekulSpiralOrder2021,zaletelBultinckGroundStateHidden2020}. This technique has been extended to bulk \tapdte{} \cite{hasanHossainDiscoveryTopologicalExciton2023} and WTe$_2$ \cite{papajPapajSpectroscopicSignaturesExcitonic2023,Watson2025WTe2}, putative exciton insulators, to detect signatures of finite-momentum interband coherence. 

\begin{figure}[htpb]
    \centering
    \includegraphics[width=\columnwidth]{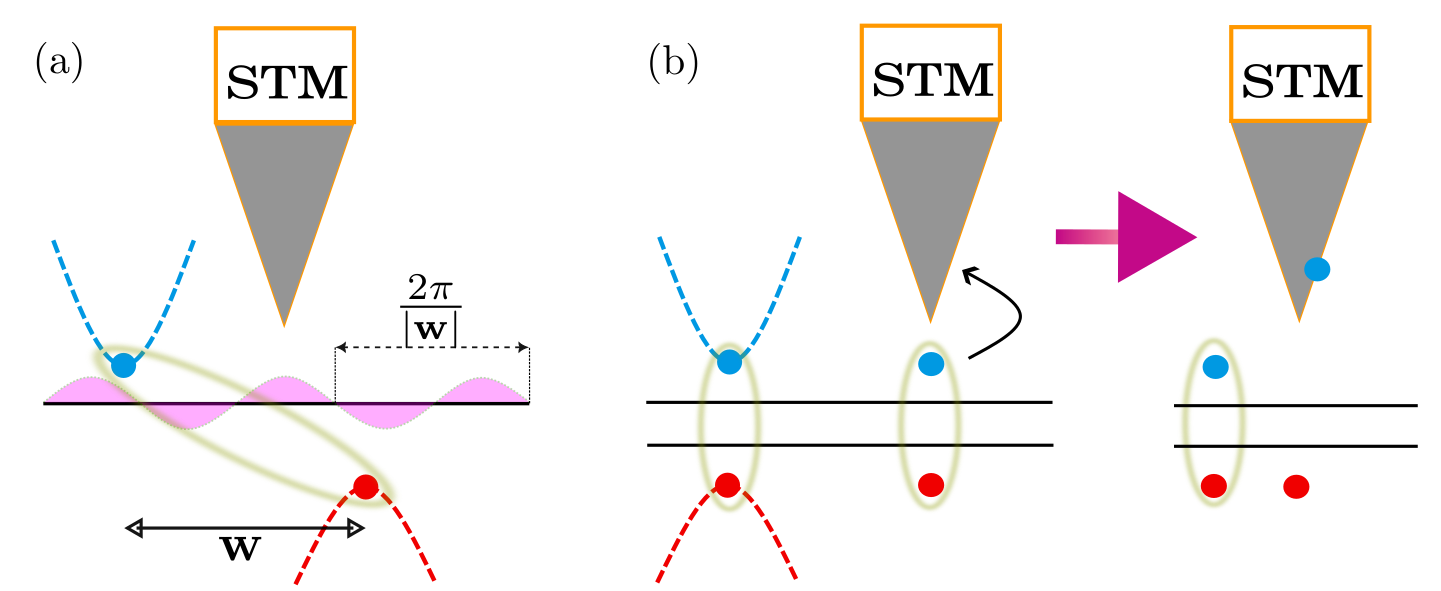}
    \caption{
(a) Schematic rendering of STM measurements on a monolayer exciton insulator. The tip tunnel couples to both bands. A momentum offset between the band edges of the conduction and valence bands by a wavevector $\mathbf w$ leads to spatial oscillations in the STM signal with wavelength ${2\pi}/{|\mathbf w|}$. These are a direct signature of interband coherence. Moreover, in combination with the spatially uniform contribution to tunneling, these allow the exciton wavefunction to be extracted. (b) Schematic rendering of a STM experiment on a bilayer exciton condensate. The tip electrons are tunnel coupled only to the top layer. Exciton condensation enables a process in which a conduction electron bound in an exciton tunnels out into the tip, leaving behind a valence hole. This leads to a characteristic peak in the tunneling spectrum, indicative of the formation of excitons.}
  \label{fig:figone}
\end{figure}

In this work, we theoretically study how STM can be applied to excitonic insulators, going beyond the paradigm of establishing a gapped state. To emphasize this point, we focus on the BEC regime. In this regime, there is a gap even in the absence of exciton formation. In fact, the regime of low exciton density has two advantages compared to the BCS regime. First, the effect of exciton-lattice coupling is smaller \cite{riceHalperinExcitonicStateSemiconductorSemimetal1968}, rendering a competing lattice-driven transition less likely. Second, screening effects are negligible for small electron-hole density. In contrast, screening is crucial in the BCS regime, particularly in bilayer devices, where it causes an early transition into an electron-hole plasma as the density increases \cite{needsMaezonoExcitonsBiexcitonsSymmetric2013,senatoreDePaloExcitonicCondensationSymmetric2002}. Moreover, the BEC regime allows for a systematic theoretical approach by solving the mean-field equations in powers of the exciton density.

In monolayer samples (Fig~\ref{fig:figone}a), in which excitonic coherence forms between conduction and valence bands offset by a certain momentum (so that the band gap is indirect), one expects lattice-symmetry breaking spatial oscillations in the tunneling signal. Depending on the ordering, these occur in the spin-averaged or the spin-resolved tunneling. These oscillations reflect that the excitons are forming a coherent condensate. Importantly, we show that combining measurements of the spatially averaged and oscillatory contributions to the tunneling conductance allows the exciton wavefunction to be extracted. Furthermore, we predict finite peaks at the band edge in the spatially averaged tunneling conductance. This contrasts with the divergence at the gap edge found in the BCS regime. Our analytical expressions for the peaks and their shape predict sharper peaks on the side of the heavier band.

In a bilayer sample, the tip electrons only couple directly to the top layer, as shown Fig.~\ref{fig:figone}b. Assuming that the conduction band is in the top layer and the valence band in the bottom layer, there is no tunneling out of the top layer in the absence of interlayer correlations. However, such tunneling processes become possible in the exciton-insulator state, as depicted in Fig.~\ref{fig:figone}b. In the exciton insulator, an electron in the conduction band of the top layer bound into an exciton can tunnel into the tip, leaving behind a hole in the bottom layer. This process is only possible if excitons are present, and leads to the emergence of a characteristic peak in the tunneling conductance upon formation of the exciton condensate. The strength of this peak can be used to extract the exciton density.

Our manuscript is structured as follows. Section \ref{sec:hf}  reviews the basic model of an exciton insulator, including a solution using mean-field decoupling and a low-density expansion. In Sec.\ \ref{sec:tunnelingprobes}, we formulate the framework for computing the local tunneling conductance as measured in scanning tunneling microscopy measurements. In Sec.\ \ref{sec:monolayer}, we study the tunneling signatures in a minimal model of a spinless monolayer exciton insulator, focusing on signatures of interband coherence and a method to recover the exciton wavefunction. Section \ref{sec:bilayer} discusses signatures of exciton formation in bilayer condensates. In Sec.\ \ref{sec:spinful}, we go beyond the minimal model and discuss how our results can be extended to include the effects of spin and valley degrees of freedom. Finally, we conclude
in Sec.~\ref{sec:discussion}.

\section{Exciton condensates}
\label{sec:hf}
\subsection{Basic model} 

We begin our discussion with an elementary two-band model of exciton condensation \cite{riceHalperinExcitonicStateSemiconductorSemimetal1968,nozieresComteExcitonBoseCondensation1982}, which neglects the spin degree of freedom. It describes the  conduction and valence bands within the effective-mass approximation,  
\begin{eqnarray}
\label{eq:spenergies}
    \epsilon^\cond_{\mathbf{k}} &=& \frac{\hbar^2 \mathbf{k}^2}{2m_c} + \frac{1}{2}E_G \\
\epsilon^\val_{\mathbf{k}} &=& -\frac{\hbar^2 \mathbf{k}^2}{2m_v}-\frac{1}{2}E_G.
\end{eqnarray}
We assume that the bands are isotropic with effective mass 
$m_c$ ($m_v$) of the conduction (valence) band and separated by a band gap $E_G$. The momenta $\mathbf{k}$ are measured from the respective band extrema and we assume charge neutrality, with equal numbers of conduction-band electrons and valence-band holes. The corresponding Bloch functions are
\begin{equation}
   \varphi_{\mathbf{k},i}(\mathbf{r}) =  \frac{e^{i(\mathbf k + \mathbf w_i)\cdot \mathbf r}}{\sqrt{\nsites}}u_{\mathbf k+\mathbf w_i,i}(\mathbf r), 
\end{equation}
where $\nsites$ is the number of unit cells, $\mathbf w_i$ is the location of the band extremum of band $i=c,v$, and $u_{\mathbf k+\mathbf w_{i},i}(\mathbf r)$ is the periodic part of the Bloch function normalized within the unit cell. The conduction and valence band operators $c^\dagger_{\mathbf k,\cond}$ and $c^\dagger_{\mathbf k,\val}$ are given as
\begin{equation}
\label{eq:operatorexpansion}
c^\dagger_{\mathbf k,i} = \int d \mathbf r\,\varphi_{\mathbf{k},i}(\mathbf{r}) \psi^\dagger(\mathbf r),
\end{equation}
where $\psi^\dagger(\mathbf r)$ creates an electron at position $\mathbf r$. The extrema of conduction and valence band are offset in momentum by $\mathbf w = \mathbf w_c - \mathbf w_v$. With a view towards van der Waals materials, we assume the Bloch states to be perfectly localized along the $z$-direction, with band $i$ localized at $z_i$. We choose $z_\cond=0$, so that $z_\cond = z_\val = 0$ for monolayers and $z_v<0$ for a bilayer. For this choice, the conduction-band layer is above the valence-band layer as in Fig.~\ref{fig:figone}b. 

Including the Coulomb interaction, the Hamiltonian takes the form
\begin{align}
    & H = \sum_{\mathbf{k}}
    \epsilon^\cond_{\mathbf{k}}  \condop^\dagger \condop^{\phantom{\dagger}} +
     \sum_{\mathbf{k}}\epsilon^\val_{\mathbf{k}} \valop^\dagger 
    \valop^{\phantom{\dagger}}  
    \nonumber\\
    & + \frac{1}{2A} \sum_{\mathbf{q}} : V^{\cond \cond}_\mathbf{q}  \rho^\cond_\mathbf{q} 
    \rho^\cond_{-\mathbf{q}} +
      V^{\cond \val}_\mathbf{q}  \rho^\cond_\mathbf{q} 
    \rho^\val_{-\mathbf{q}} + (\cond \leftrightarrow \val): + E_C
\label{eq:hamsimple}
\end{align}
in terms of the electron density $\rho^{\cond/\val}_{\mathbf q}$ in the conduction/valence band. Here, $A$ is the sample area, normal ordering is with respect to the intrinsic semiconductor (i.e., conduction-band electrons and valence-band holes), and $V^{ij}_{\mathbf q}$ is the unscreened Coulomb interaction for bands $i,j$,
\begin{equation}
\label{eq:vcoulomblayereddef}
V^{ij}_{\mathbf q }=  \frac{e^2}{2\epsilon \epsilon_0|\mathbf q|} \exp(-|\mathbf q|\cdot|z_i-z_j|),
\end{equation}
with the electron charge $e$, vacuum permitivity $\epsilon_0$, and a background relative dielectric constant $\epsilon$. We included the electrostatic charging energy $E_C$ of a parallel-plate capacitor
\begin{equation}
\label{eq:capacitance}
    E_C = 
\frac{1}{2A C_G}:\Nx^2:,
\end{equation}
with the interlayer capacitance per unit area $C_G =\frac{\epsilon \epsilon_0}{e^2} |z_\cond-z_\val|^{-1}$ and the exciton number operator $\Nx =\frac{1}{2} \sum_{\mathbf k}(\condop^\dagger\condop^{\phantom{\dagger}}-\valop^{\phantom\dagger} \valop^{\dagger})$. The charging energy $E_C$ vanishes for a monolayer condensate. 

The Hamiltonian in Eq.\ \eqref{eq:hamsimple} assumes the dominant-term approximation, which neglects the momentum dependence of wavefunction overlaps and interband scattering \cite{riceHalperinExcitonicStateSemiconductorSemimetal1968}. In this approximation, the electron density is just the sum of the densities 
\begin{equation}
\label{eq:defrhodta}
\rho^i_{\mathbf q}  = \sum_{\mathbf k} c^{\dagger}_{\mathbf k,i}c^{\phantom{\dagger}}_{\mathbf k +\mathbf q,i}
\end{equation}
in the two bands. Neglecting the momentum dependence of band overlaps is equivalent to approximating $u_{\mathbf k+\mathbf w_i,i}(\mathbf r) \simeq u_{\mathbf w_i,i}(\mathbf r)$. This is justified when $u_{\mathbf k+\mathbf w_i,i}(\mathbf r)$ is slowly varying with $\mathbf k$ on the scale of the inverse Bohr radius of the excitons. In what follows, we also use the simplified notation $u_{\mathbf w_i,i}(\mathbf r) \to u_i(\mathbf r)$. In general, the electron density involves mixed terms involving both conduction and valence band operators, leading to interband scattering. For monolayer systems, these are suppressed when conduction and valence bands are offset by a finite momentum $\mathbf w$ (Fig.\ \ref{fig:figone}a) and the Coulomb potential $V^{}_{\mathbf w}\propto 1/|\mathbf w|$ is small relative to intraband contributions \cite{riceHalperinExcitonicStateSemiconductorSemimetal1968}. For bilayer systems with conduction and valence bands residing in different layers, the neglect of interband scattering is justified by the vanishing overlap between the conduction and valence band wavefunctions \cite{riceZhuExcitonCondensateSemiconductor1995,novoselovFoglerHightemperatureSuperfluidityIndirect2014,yakobsonGuptaHeterobilayers2DMaterials2020} (Fig.~\ref{fig:figone}b).

Within the dominant-term approximation, interband scattering is neglected and the model has separate charge-conservation symmetries for conduction and valence electrons. Interband coherence spontaneously breaks these separate symmetries, leaving only overall charge conservation. In monolayer systems, interband coherence results in spatial density modulations at the  wavevector $\mathbf w$, forming a charge density wave (CDW) within the spinless model. The origin of this CDW is arbitrary, leading to a gapless Goldstone mode. However, interband scattering (neglected in the dominant-term approximation) explicitly breaks the separate charge conservation symmetries and generally pins the origin of the CDW, gapping out the Goldstone mode. Consequently, while monolayer exciton condensates spontaneously break the lattice translation symmetry, 
the independent charge conservation symmetries are explicitly broken by interband scattering terms in the interaction.

\subsection{Mean-field solution} 
\label{subsec:meanfieldpert}

To set the stage, we review the mean-field solution, focusing on two dimensions. The Coulomb attraction between electrons in the conduction band and holes in the valence leads to spontaneous condensation of electron-hole pairs (excitons) as the band gap $E_G$ is reduced below the exciton binding energy $E_b$ \cite{riceHalperinExcitonicStateSemiconductorSemimetal1968,maksimovKozlovMetalDielectricDivalentCrystal1965,kozlovKeldyshCollectivePropertiesExcitons1968}. In the BEC limit, we can view $\sum_{\mathbf k} \phi_{\mathbf k}c^{\dagger}_{\mathbf k,\cond}c^{\phantom{\dagger}}_{\mathbf k,\val}$ as the creation operator of an exciton in the hydrogenic 1s state, where $\phi_{\mathbf k}$ denotes the corresponding wave function in momentum representation. We use the normalization $({1}/{A})\sum_{\mathbf k}\phi^2_{\mathbf k}=1$. The corresponding Bose-condensed ground state takes the form \cite{nozieresComteExcitonBoseCondensation1982}
\begin{equation}
 \ket{\Psi_\lambda} = \exp\left(\lambda \sum_{\mathbf k}  \phi_{\mathbf k}c^{\dagger}_{\mathbf k,\cond}c^{\phantom{\dagger}}_{\mathbf k,\val}\right) \ehvacuum,
 \label{eq:xcohst}
\end{equation}
where $\ehvacuum = \prod_\mathbf{k} c^\dagger_{\mathbf{k},v}\vacuum$ is the ground state of the intrinsic semiconductor. The eigenvalue $\lambda$ of the coherent state controls the exciton density, with the exciton density $n_\mathrm{ex} \simeq |\lambda|^2$ in the dilute limit. 

More generally, the exciton wave function $\phi_\mathbf{k}$ can be viewed as a variational parameter
\cite{nozieresComteExcitonBoseCondensation1982}. Expanding the exponential, the (normalized) coherent state in Eq.\ \eqref{eq:xcohst} can be rewritten as 
\begin{equation}
\ket{\Psi_\lambda} = \Pi_{\mathbf{k}}(
        \uk c^{\dagger}_{\mathbf k,\val} + \vk c^{\dagger}_{\mathbf k,\cond}     
        )\vacuum,
        \label{eq:trial}
\end{equation}
where 
\begin{equation}
u_{\mathbf k} = \frac{1}{\sqrt{1+\lambda^2 \phi_{\mathbf k}^2}} \quad ; \quad v_{\mathbf k} = \frac{\lambda \phi_{\mathbf k}}{\sqrt{1+\lambda^2 \phi_{\mathbf k}^2}}.
\end{equation}
Thus, the BEC ansatz is in fact a Hartree-Fock approximation in terms of the variational Slater determinant $\ket{\Psi_\lambda}$, which is not limited to the BEC limit.

The optimal Slater determinant is obtained by mean-field decoupling the full interacting Hamiltonian of Eq.\  \eqref{eq:hamsimple}. This leads to the mean-field Hamiltonian
\cite{riceHalperinExcitonicStateSemiconductorSemimetal1968,nozieresComteExcitonBoseCondensation1982,riceZhuExcitonCondensateSemiconductor1995}
\begin{equation}
    \label{eq:HBdG}
    {H}_{\text{MF}} = \sum_{\mathbf{k},\sigma}  
\begin{pmatrix}
\condop^\dagger &
\valop^\dagger
\end{pmatrix}
h_\mathrm{mf}
\begin{pmatrix}\condop^{\phantom{\dagger}}\\
\valop^{\phantom{\dagger}} \end{pmatrix} \,\, ; \,\,
h_\mathrm{mf} = \begin{pmatrix} \bar{\epsilon}^\cond_{\mathbf{k}}& \Delta_{\mathbf{k}} \\ 
\Delta_{\mathbf{k}} & \bar\epsilon^\val_{\mathbf{k}}\end{pmatrix}.
\end{equation}
Here, the single-particle energies 
\begin{eqnarray}
\overline{\epsilon}^\cond_\mathbf{k}
& = & \epsilon^\cond_\mathbf{k} - \frac{1}{A}\sum_{\mathbf{k}'}V^{\cond \cond}_{\mathbf{k}-\mathbf{k}^\prime} \langle \condopp^\dagger \condopp^{\phantom{\dagger}} 
\rangle 
+\frac{\nx}{2C_G}, 
   \\ 
\overline{\epsilon}^\val_\mathbf{k} &=&  \epsilon^\val_\mathbf{k} + \frac{1}{A}\sum_{\mathbf{k}^\prime}V^{\val \val}_{\mathbf{k}-\mathbf{k}^\prime} \langle \valopp^{\phantom{\dagger}} \valopp^{{\dagger}} \rangle 
 -\frac{\nx}{2C_G} 
\end{eqnarray}
are renormalized by Fock and (for bilayer systems) interlayer capacitance terms. Moreover, $\nx = \langle \Nx \rangle/A$ denotes the exciton density and 
\begin{eqnarray}
\Delta_{\mathbf k} & = & -\frac{1}{A}\sum_{\mathbf k} V^{\cond \val}_{\mathbf k- \mathbf k'} \langle \condopp^\dagger \valopp^{\phantom{\dagger}}  \rangle.
\end{eqnarray}
measures the strength of interband coherence. We choose the order parameter to be real, which can be achieved by a suitable gauge transformation. Above, $\langle O \rangle$ denotes the expectation value of an operator $O$ in the trial state in Eq.\ \eqref{eq:trial}. The self-consistent ground state of Eq.~\eqref{eq:HBdG} is the Slater determinant, which minimizes the expectation value of the full Hamiltonian \cite{fockFockNaeherungsmethodeZurLoesung1930}.

According to Koopmans' theorem, the mean-field Hamiltonian $h_\mathrm{mf}$ in Eq.\ \eqref{eq:HBdG} describes the 
single-particle excitations \cite{koopmansKoopmansUberZuordnungWellenfunktionen1934}. This yields
two bands with energies 
\begin{eqnarray}
\label{eq:bandenergiesdef}
{E_{\mathbf{k},\pm} = \meanenergy \pm \bcsenergy}.
\end{eqnarray}
and eigenvectors $(\uk, -\vk)$ for the upper ($+$) band and $(\vk, \uk)$ for the lower ($-$) band. The corresponding Bogoliubov operators are 
\begin{align}
  d_{\mathbf{k},+}^{\dagger}&= u_{\mathbf{k}}  
\condop^{\dagger}
-\vk \valop^{\dagger},\\
d_{\mathbf{k},-}^{\dagger} &= v_{\mathbf{k}} 
     \condop^{\dagger}+
\uk \valop^{\dagger}.
\label{eq:Bogodintermsofc}
\end{align}
In terms of these operators, the mean-field Hamiltonian in Eq.\ \eqref{eq:HBdG} is diagonal, $ H_\mathrm{MF} = \sum_{\mathbf k,\alpha=\pm} E_{\mathbf k,\alpha}d_{\mathbf{k},\alpha}^{\dagger}d_{\mathbf{k},\alpha}^{\phantom{\dagger}}$.

At charge neutrality, all 
$d^\dagger_{\mathbf k,-}$ orbitals are occupied, while
all $d^\dagger_{\mathbf k,+}$ orbitals are empty. Thus, the mean-field ground state is simply
\begin{align}
\label{eq:excitongs}
 \GS &=
\Pi_{\mathbf{k}}( \uk\valop^\dagger + v_{\mathbf{k}}  \condop^{\dagger}     
        )\vacuum  
        \nonumber\\
        &=
\Pi_{\mathbf{k}}(
        \uk +\vk \condop^{\dagger}    \valop^{\phantom{\dagger}}  )\ehvacuum  ,
        \end{align}
where the first line emphasizes the Hartree-Fock nature and the second the electron-hole pairing of $\GS$.
Using the expectation values $\langle \condop^\dagger \condop^{\phantom{\dagger}} 
\rangle = \langle \valop^{\phantom{\dagger}} \valop^{\dagger} 
\rangle = \vk^2$ and
$\langle \condop^\dagger \valop^{\phantom{\dagger}} 
\rangle = \uk \vk$ 
in this state, the self-consistency equation reads
\begin{multline}
\label{eq:selfconsistencyeq}
\left(\frac{\hbar^2 \mathbf k ^2}{2m}+E_g +\frac{1}{A}\sum_{\mathbf k}\frac{\vk^2}{C_G}-\frac{2}{A}\sum_{\mathbf k'}V^{cc}_{\mathbf k-\mathbf k'}v_{\mathbf k'}^2\right) \uk \vk \\
=\left[\frac{1}{A}\sum_{\mathbf k'}V^{cv}_{\mathbf k-\mathbf k'}u_{\mathbf k'}v_{\mathbf k'}\right](\uk^2-\vk^2),
\end{multline}
with the reduced mass $m = m_c m_v/(m_c+m_v)$. 

For both monolayers and bilayers, we will focus on the low-exciton-density (BEC) regime, where we can exploit a systematic expansion of the mean-field solution in powers of the exciton density \cite{kozlovKeldyshCollectivePropertiesExcitons1968,nozieresComteExcitonBoseCondensation1982}.
For $E_G>E_b$, there is only the trivial solution $\uk=1, \vk= 0$. For $E_G<E_b$, there will be a finite density of excitons in the ground state given by $\nx = ({1}/{A})\sum_{\mathbf k}\vk^2$. To lowest order in the low-density limit (small $\vk$), Eq.~\eqref{eq:selfconsistencyeq} reduces to the excitonic Schr\"odinger equation \cite{nozieresComteExcitonBoseCondensation1982} 
\begin{equation}
\label{eq:momentumspaceexcitonschroedinger}
\frac{\hbar^2k^2}{2m} \vk - \frac{1}{A}\sum_{\mathbf k'} V^{\cond \val}_{\mathbf k- \mathbf k'} v_{\mathbf k'} =- E_G\vk.
\end{equation}
This equation is solved by setting $E_G$ equal to the binding energy $E_b$ of the hydrogenic 1s wavefunction $\vk = \lambda \phi_{\mathbf k}$. At this order, one recovers the coherent state in Eq.~\eqref{eq:xcohst}. In the monolayer case with ${1}/{C_G}=0$ and a 2D Coulomb interaction, Eq.~\eqref{eq:momentumspaceexcitonschroedinger} can be solved analytically, giving a binding energy $E_b = 4\rydberg$ and the wavefunction
\begin{equation}
\label{eq:excitonwfmonolayer}
\phi_{\mathbf k} =\frac{\sqrt{2\pi} \bohrradius }{ \left[ 1+(k\bohrradius)^2/4 \right]^{\frac{3}{2}}}
\end{equation}
in terms of the excitonic Rydberg constant ($\rydberg$) and Bohr radius ($\bohrradius$). 

The exciton density is determined by considering higher-order contributions in Eq.~\eqref{eq:selfconsistencyeq}. We perform this derivation in App.\ \ref{app:solutionofselfconsistencyeq}. For small $\nx$, we find $E_G-E_b = \frac{d\mux}{d\nx} \nx$, where the inverse exciton compressibility $\frac{d\mux}{d\nx}$ can be expressed in terms of the bound-state exciton wavefunction $\phi_{\mathbf k}$ and the Coulomb interaction only. The exciton wavefunction $\phi_{\mathbf k}$ together with the exciton compressibility also enables the mean-field band energies $E_{\mathbf{k},\pm}$ to be evaluated to linear order in $\nx$. For the monolayer wavefunction, Eq.~\eqref{eq:excitonwfmonolayer}, the intraband Fock and pairing terms can be evaluated analytically, while the integral entering the numerical prefactor of $\frac{d\mux}{d\nx}$ has to be evaluated numerically.

\section{Tunneling conductance}
\label{sec:tunnelingprobes}

We compute the tunneling conductance $\didv$  between tip and sample within Fermi's golden rule. It is convenient to express the conductance in terms of 
\begin{equation}
\label{eq:gnormal}
G_0 = \frac{2\pi e^2}{\hbar}g_s t^2 \nu_0 \nu_t,
\end{equation}
which corresponds to a spatially averaged conductance into a parabolic band (density of states $\nu_0 = m/2\pi\hbar^2$), whose mass is equal to the reduced mass $m$. Here, $t$ denotes the tunneling amplitude between tip (density of states per spin $\nu_t$) and sample, while $g_s=2$ is a spin degeneracy factor. Apart from the replacement $m \leftrightarrow m_\cond$  ($m_\val$), this corresponds to the normal-state conductance into the conduction (valence) band. Denoting the electron operator at the position $\rtip$ of the tip as $\psi(\rtip)$, the zero-temperature tunneling conductance at bias voltage $V$ can be expressed as  
\begin{multline}
\label{eq:didvstatei}
\didv = G_0 \frac{1}{\nu_0} \sum_{f}\Big[ |\bra{f}\psi^\dagger(\rtip) \GS |^2 \delta(E_{f}-E_0-eV) \\
+|\bra{f}\psi(\rtip) \GS |^2 \delta(E_0-E_f-eV)\Big].
\end{multline}
The initial state corresponds to the many-body ground state $\ket{\mathrm{GS}}$ with energy $E_0$ and we sum over all many-body final states $\ket{f}$ with energy $E_f$. Here, the first (second) term describes tunneling into (out of) the sample. We remark that for multilayer samples, the tip position $\rtip$ also specifies the layer, which electrons are tunneling into.

Equation \eqref{eq:didvstatei} simplifies considerably for a fermionic mean-field system. The electron operator $\psi(\rtip)$ can be expanded in the band operators by inverting Eq.~\eqref{eq:operatorexpansion}, 
\begin{equation}
\psi^\dagger(\rtip) = \sum_{\mathbf k}\sum_{i}
\varphi_{\mathbf{k},i}^*(\rtip) c^\dagger_{\mathbf k,i}.
\end{equation}
We can further write this in terms of the Bogoliubov operators by inverting Eq.\ \eqref{eq:Bogodintermsofc}. This gives
\begin{equation}
\psi^\dagger(\rtip) = \sum_{\mathbf k} \sum_{\alpha = \pm}
\varphi_{\mathbf{k}}^\dagger(\rtip) \cdot \mathbf{e}_{\mathbf{k,\alpha}} d^\dagger_{\mathbf k,\alpha}.
\end{equation}
Here, we write the Bloch functions as spinors in band space, $\varphi_\mathbf{k}^T = (\varphi_{\mathbf{k},c}, \varphi_{\mathbf{k},v})$ and define the unit vectors
\begin{equation}
   \mathbf{e}_{\mathbf{k,+}}
   = \left(\begin{array}{c} u_\mathbf{k} \\ -v_\mathbf{k}\end{array}\right)
   \quad ; \quad
   \mathbf{e}_{\mathbf{k,-}} = \left(\begin{array}{c} v_\mathbf{k} \\ u_\mathbf{k}\end{array}\right).
\end{equation}
Since all $d^\dagger_{\mathbf k,-}$ orbitals are occupied in the ground state, the sum over final states is a sum over the states $\ket{f_{\mathbf{k},+}} = d^\dagger_{\mathbf k,+}\ket{\mathrm{GS}}$ for tunneling into the sample and $\ket{f_{\mathbf{k},-}} = d_{\mathbf k,-}\ket{\mathrm{GS}}$ for tunneling out of the sample. The matrix-element factors can then be written as
\begin{align}
  |\bra{f_{\mathbf{k},+}}\psi^\dagger(\rtip) \GS |^2 
 &= \Tr \{\Phi(\mathbf{k},\rtip)P_+ (\mathbf{k})\}
 \\
 |\bra{f_{\mathbf{k},-}}\psi(\rtip) \GS |^2 
 &= \Tr \{\Phi(\mathbf{k},\rtip) P_-(\mathbf{k}) \}
\end{align}
in terms of the Bloch-function factor 
\begin{equation}
   [\Phi(\mathbf{k},\rtip)]_{i,j} = \varphi_{\mathbf{k},i}(\rtip) \varphi_{\mathbf{k},j}^*(\rtip) 
   \label{eq:spinlessPhi}
\end{equation}
and the projectors
\begin{equation}
    [P_\alpha(\mathbf{k})]_{i,j} = (\mathbf{e}_{\mathbf{k,\alpha}})_i (\mathbf{e}_{\mathbf{k,\alpha}})_j^*
\end{equation}
characterizing the Bogoliubov excitations of the exciton insulator. We further neglect the weak dependence of the Bloch functions on $\mathbf{k}$, so that $\Phi(\mathbf{k},\rtip)$ takes the explicit form
\begin{equation}
\wfvector_{i,j}(\rtip) \simeq
 \frac{1}{N_\mathrm{UC}} e^{i(\mathbf w_i-\mathbf w_j)\cdot \rtip} u_{\mathbf w_i,i}(\rtip)u^*_{\mathbf w_j,j}(\rtip).
\label{eq:wfvectorspinless} 
\end{equation}
The tunneling conductance can now be compactly written as  
\begin{equation}
\label{eq:didvstatepexprssion}
\didv = G_0 \frac{1}{\nu_0} \sum_{\mathbf k, \indexsolution} \Tr\left[\wfvector(\rtip) 
 P_\alpha(\mathbf k)\right]
\deltazero .
\end{equation}
This expression separates the real-space dependence entering through the band wavefunctions and thus $\wfvector(\rtip)$
from the mean-field solution encoded by $P_\alpha$. In fact, the latter 
is independent of the detailed wavefunctions when working within the dominant-term approximation.

\begin{figure}[t]
    \centering
    \includegraphics[width=0.95\columnwidth]{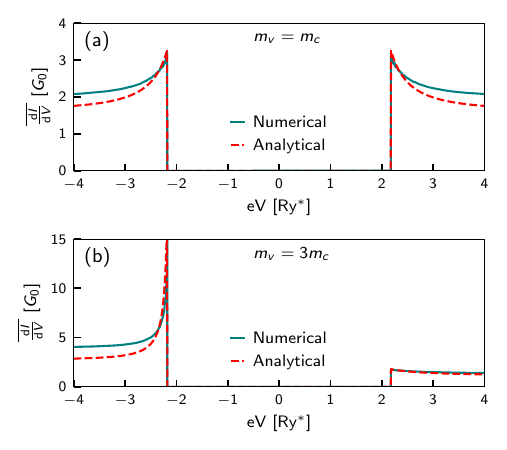}
    \caption{
Spatially averaged tunneling conductance for a monolayer, see Eq.\ \eqref{eq:spatiallyaveragedspinlessdos}. Numerical results (blue) are compared to analytical results from Eqs.\ \eqref{eq:analyticaldosfinalupper} and \eqref{eq:analyticaldosfinallower}, working at $\nx=0.025\, (\bohrradius)^{-2}$. The panels show results for (a) equal masses, $m_c=m_v$ and (b)  unequal masses, $m_v=3m_c$.}
    \label{fig:figtwo}
\end{figure}

\section{Monolayer condensates}
\label{sec:monolayer}

For tunneling into a monolayer condensate, both the diagonal and the offdiagonal components of the Bloch-function factor $\wfvector(\rtip)$ contribute. While the diagonal components reflect the lattice periodicity, the offdiagonal components  oscillate spatially with wavevector $ \mathbf w$ and do not respect the lattice symmetry except in systems with a direct gap. Thus, the offdiagonal contributions reflect the translation-symmetry breaking associated with the interband coherence of the exciton condensate. We can gain insight into the offdiagonal contributions by considering, say, tunneling into the sample and writing 
\begin{equation}
  d_{\mathbf{k},+}^{\dagger}= u_{\mathbf{k}} \condop^{\dagger}
-\vk h_{-\mathbf{k},v} 
\end{equation}
in terms of the hole operator $h_{-\mathbf{k},v} = \valop^{\dagger}$. This implies that the amplitude for tunneling into the sample involves two processes. One process adds an unbound electron to the conduction band, while the other breaks up an exciton by removing a valence-band hole, in effect also creating an unbound conduction-band electron. Both processes contribute coherently due to the condensate formation, with the interference term just corresponding to the contribution of the offdiagonal components of the Bloch-function factor $\wfvector(\rtip)$. 

The interference terms do not contribute to the spatially averaged conductance. Noting that $\int d\rtip \wfvector(\rtip) = \mathbb{1}$, one obtains 
\begin{equation}
\label{eq:spatiallyaveragedspinlessdos}
\overline{\didv} = G_0 \frac{1}{\nu_0 A} \sum_{\mathbf k, \indexsolution} \deltazero ,
\end{equation}
where the overline denotes the spatial average. Thus, the spatially averaged conductance measures the total density of states of the mean-field spectrum. 

\subsection{Spatially averaged tunneling conductance}
\label{subsec:monolayeraverage}

We can use the framework of Sec.\ \ref{subsec:meanfieldpert} to find analytical expressions for the spatially averaged tunneling conductance in the BEC limit. Using the analytical exciton wavefunction in Eq.~\eqref{eq:excitonwfmonolayer} and working to linear order in $\nx$, we find that exciton condensation changes the excitation gap to
\begin{equation}
E_{0,+}-E_{0,-} =\left[4-6.0\nxatomic +16\pi \nxatomic- 3\pi^2\nxatomic\right]\, \rydberg.
\end{equation}
Here, we write the exciton density in atomic units as $\nxatomic = \nx (\bohrradius)^{2}$. The first two terms in the square brackets are the single-particle band gap $E_G= E_b -\frac{d \mux}{d \nx}\nx $, with monolayer binding energy $E_b=4\,\rydberg$ and inverse compressibility $\frac{d \mux}{d \nx}= 6.0\,\rydberg (\bohrradius)^{2}$. The third term accounts for the increase of the interacting gap due to the offdiagonal pairing term $\Delta_{\mathbf k}$ in Eq.~\eqref{eq:HBdG}. The last term is the intraband Fock term evaluated at $\mathbf k=0$, which reduces the gap. In total, the formation of the exciton condensate increases the excitation gap.

Using the same expansion in the limit of small exciton densities, we can also obtain approximate expressions for $\didv$ close to the tunneling thresholds $eV = E_{0,\pm}$. We find 
\begin{equation}
\label{eq:analyticaldosfinalupper}
\overline \didv \simeq \frac{G_0 \Theta(eV - E_{0,+})}
{\frac{m}{m_c}+\frac{15 \pi^2}{64} \nxatomic
- 
4\pi \nxatomic
 \left[1+\frac{m_c}{m}\frac{eV-E_{0,+}}{E_b} \right]^{-3} }
\end{equation}
for positive bias ($eV>0$) and 
\begin{equation}
\label{eq:analyticaldosfinallower}
\overline \didv \simeq \frac{G_0 \Theta( E_{0,-} - eV)} 
{ 
\frac{m}{m_v}+\frac{15 \pi^2}{64} \nxatomic
- 4\pi \nxatomic
 \left[1+\frac{m_v}{m}\frac{E_{0,-}-eV}{E_b}\right]^{-3}  }
\end{equation}
for negative bias ($eV<0$). In these expressions, the first term in the denominators is the single-particle contribution, the second term is a Fock mass renormalization, and the third term is due to the excitonic pairing. The last term results in a tunneling peak at the band edge. For differing conduction and valence band masses, this peak will be more pronounced on the side of the heavier band. 

A numerical solution of the mean-field equations confirms these analytical expectations. Figure \ref{fig:figtwo} compares the analytical results (shown in red) to numerical Hartree-Fock results (shown in blue, see App.\ \ref{app:numerical} for details) for equal (Fig.\ \ref{fig:figtwo}a) and different (Fig.\ \ref{fig:figtwo}b) band masses. We find good agreement  close to the band edges. In particular, the analytical results accurately capture the more pronounced effect of exciton condensation on the side of the heavier band. 

Unlike in BCS theory, the differential conductance peaks at the gap edge do not diverge. This can be traced to the fact that here, the excitation gap is governed by the band edges and not by states at the Fermi energy away from the band edges as in the BCS limit.  

\begin{figure}[t]
    \centering
    \includegraphics[width=0.95\columnwidth]{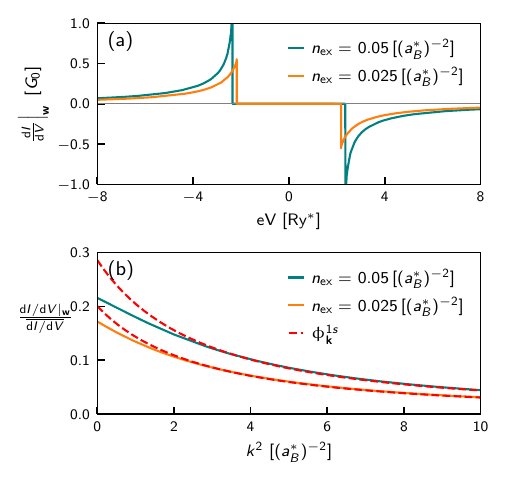}
    \caption{(a) Fourier component of differential conductance at wavevector $\mathbf w$ for $\nx (\bohrradius)^2 = 0.025 $ (orange) and $\nx (\bohrradius)^2 =  0.05$ (blue). (b) Exciton wavefunction extracted  by combining numerical results for the spatially oscillating and averaged differential conductance, plotted as a function of $k^2 = \frac{e \overline I}{G_0\,\rydberg} (\bohrradius)^{-2}$, cf.\  Eq.~\eqref{eq:currentkconversion}. The dashed red line is the analytical wavefunction, Eq.~\eqref{eq:excitonwfmonolayer}, for comparison. Only a constant prefactor is fitted. Both panels: $m_c=m_v$. }
    \label{fig:figthree}
\end{figure}

\subsection{Spatial oscillations of the tunneling conductance}
\label{subsec:monolayerosc}

As discussed above, the local tunneling conductance exhibits spatial oscillations with the wavevector $\mathbf{w}$ due to the formation of the exciton condensate. Thus, a spatial Fourier transform of $\didv$ has Fourier components at wavevectors $\mathbf w + \mathbf G$, where $\mathbf G$ denotes the reciprocal lattice vectors. We find the explicit expressions
\begin{multline}
\label{eq:didvstatepexprssionmonolayerw}
\left.\didv\right|_{\mathbf w + \mathbf G} = G_0\frac{1}{\nu_0} \braket{{u_{\mathbf w_\val, \val}|u_{\mathbf w_\cond+\mathbf G, \cond}}} 
\\ \times
\frac{1}{A}\sum_{\mathbf k}
\uk \vk 
 \left[\deltazero[-] - \deltazero[+]\right],
\end{multline}
Typically, the wavefunction overlap $\braket{{u_{\mathbf w_\val, \val}|u_{\mathbf w_\cond+\mathbf G, \cond}}} = \int_{\text{UC}} d \mathbf r\, u^*_{\val}(\mathbf r)e^{i \mathbf G \cdot \mathbf r}u_{\cond}(\mathbf r) $ decreases rapidly with $\mathbf G$, implying the strongest contributions at wavevector $\mathbf w$. A crucial difference from the result for the spatially averaged conductance in Eq.\ \eqref{eq:spatiallyaveragedspinlessdos} is the additional weight factor $\uk \vk $. Remembering that $\langle \condop^\dagger \valop^{\phantom{\dagger}} 
\rangle = \uk \vk$, the spatial oscillations of the tunneling conductance are a direct signature of interband coherence. Thus, these lattice-symmetry breaking oscillations in $\didv$ are a signature of a coherent exciton condensate and would not be present for a putative incoherent exciton gas at elevated temperatures \cite{novoselovFoglerHightemperatureSuperfluidityIndirect2014}. We also remark that the oscillations are analogous to the Kekul\'e pattern due to intervalley coherence observed recently in graphene devices \cite{yazdaniLiuVisualizingBrokenSymmetry2021,nadj-pergeKimImagingIntervalleyCoherent2023,yazdaniNuckollsQuantumTexturesManybody2023}. 

We plot the oscillatory component in Fig.\ \ref{fig:figthree}a for two representative values of the exciton density $\nx$, assuming a band overlap of $\braket{u_{\mathbf w_\cond, \cond}|u_{\mathbf w_\val, \val}} = \frac{1}{2}$.
Due to the $\uk\vk$ factor, the oscillatory signal decays away from the excitation threshold on a scale given by the exciton binding energy $E_b$.

\subsection{Extracting the exciton wavefunction}
\label{subsec:monolayerrecovery}

In the limit of low exciton density, the perturbative solution of the mean-field equations gives
\begin{equation}
\label{eq:ukvkisphi}
\frac{1}{A}\uk \vk  \approx \sqrt{\nx} \phi^{1s}_{\mathbf k},
\end{equation}
with the exciton wavefunction $\phi^{1s}_{\mathbf k}$ given in Eq.\ \eqref{eq:excitonwfmonolayer}. Remarkably, this implies that measurements of the spatially averaged and spatially oscillating tunnel conductances can be combined to reconstruct the exciton wavefunction. 

We observe that the voltage bias fixes the modulus $k$ of the wavevector contributing to $\didv$. Thus, we can extract the exciton wave function by dividing the amplitude of the spatial oscillations and the spatially averaged conductance. Working at positive bias for definiteness, we have
\begin{equation}
\label{eq:division}
\left. \frac{\mathrm{d}I/\mathrm{d}V|_\mathbf{w}}{\overline{\mathrm{d}I/\mathrm{d}V}}\right|_{V>0} \propto \Theta(eV-E_+^0) u_{k(V)}v_{k(V)},
\end{equation}
where we use Eqs.\ \eqref{eq:spatiallyaveragedspinlessdos} and \eqref{eq:didvstatepexprssionmonolayerw}. The prefactor is just given by the wavefunction overlap and thus independent of $k$. 

Using rotational symmetry, the function $k(V)$ is defined implicitly through
\begin{equation}
\label{eq:defkv}
E_{k(V),+} = eV,
\end{equation}
where we assume monotonicity of $E_{k,+}$, a natural assumption in the BEC regime. To extract the change of variables from bias to momentum in experiment, one obtains the total spatially-averaged current $\overline I(V)$ at positive bias $V$, e.g., by direct measurement or by integrating $\overline \didv$ from the conduction band bottom up to the bias $V$. The spatially averaged current is related to the momentum $k(V)$ through
\begin{equation}
\label{eq:currentkconversion}
    \overline I (V)= 
\int_{E_+^0/e}^{E_{k(V),+}/e} dV \overline \didv = 
    G_0\frac{\rydberg}{e} [k(V) \bohrradius]^2.
\end{equation}
Using Eq.\ \eqref{eq:ukvkisphi}, we can therefore extract the exciton wavefunction.

We apply this procedure to a numerical Hartree-Fock solution and show corresponding results in Fig.\ \ref{fig:figthree}b. Numerical results for two different exciton densities (blue and orange lines in Fig.\ \ref{fig:figthree}b) are compared to the analytical result for $\phi_{\mathbf k}^{1s}$ in Eq.\ \eqref{eq:excitonwfmonolayer}  (red dashed lines in Fig.\ \ref{fig:figthree}b). For small $\nx$ (orange line), the exciton wavefunction can be extracted almost perfectly from this procedure. Only the overall prefactor of the wavefunction needs to be fit. On the other hand, for larger $\nx$ (blue line), there are some deviations for small momenta. The deviations appear because $\vk$ becomes significant for small momenta $\mathbf k$, so that assuming $\uk \approx 1$ as in Eq.\ \eqref{eq:ukvkisphi} is no longer valid. We conclude that the spatial oscillations in the STM measurements signal the existence of an exciton condensate. Moreover, they also enable the exciton wavefunctions to be recovered from experimental data, provided that the crystal symmetry is such that the conduction and valence-band edges have approximate rotational symmetry.
\begin{figure}[t]
    \centering
    \includegraphics[width=0.95\columnwidth]{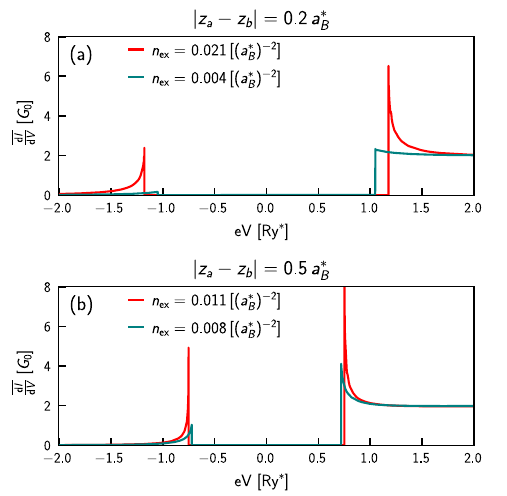}
    \caption{(a) Spatially averaged differential conductance into a bilayer exciton condensate for $|z_a-z_b|=0.2\, \bohrradius$ for two values
    of $\nx$. The appearance of a peak at negative bias is characteristic for exciton condensation. Its integrated weight provides a measure of $\nx$, cf.\ Eq.~\eqref{eq:peakintegralnxconversion}. 
    (b) Like (a) but with $|z_a-z_b|=0.5\,\bohrradius$. Both panels: $m_c=m_v$.
        }
    \label{fig:figfour}
\end{figure}

\section{Bilayer condensates}
\label{sec:bilayer}

In bilayer systems, tunneling is only possible into, say, the top layer hosting the conduction band. This can be accounted for by setting $\varphi_{\mathbf{k},v}(\rtip)=0$, so that the Bloch-function factor $[\wfvector]_{c,c} = |\varphi_{\mathbf{k},c}(\rtip)|^2$ is the only nonzero matrix element of $\wfvector(\rtip)$. As a result, unlike for monolayers the tunneling conductance 
\begin{equation}
\label{eq:didvbilayerbasic}
\didv = G_0 \frac{1}{\nu_0} \sum_{\mathbf k, \indexsolution} |\varphi_{\mathbf{k},c}(\rtip)|^2 
 [P_\alpha(\mathbf k)]_{c,c}
\deltazero ,
\end{equation}
respects the lattice symmetry. However, bilayers still allow for an interesting signature of exciton formation.  In fact, in the absence of exciton formation ($u_\mathbf{k}=1$ and $v_\mathbf{k}=0$), the tunneling conductance is nonzero for $eV>0$ only. In contrast, exciton formation leads to a nonzero tunneling conductance even for $eV<0$. In this case, only the lower quasiparticle band $\alpha = -$ contributes to the differential conductance and we find
\begin{equation}
\left.\didv\right|_{eV<0} = G_0 \frac{1}{\nu_0} \sum_{\mathbf k} |\varphi_{\mathbf{k},c}(\rtip)v_\mathbf{k}|^2
\delta(E_{\mathbf{k},-} - eV) .
\end{equation}
Physically, exciton formation implies the presence of conduction-band electrons bound to valence holes. These conduction-band electrons can tunnel out, leaving behind a hole excitation in the valence band, as illustrated in Fig.~\ref{fig:figone}a. We note that unlike the offdiagonal contribution to the tunneling conductance of monolayer condensates, this signature only reflects exciton formation, but does not require coherence of the condensate. 

Due to the absence of lattice-symmetry breaking spatial oscillations, we focus on the spatially averaged conductance
\begin{equation}
\label{eq:didvstatepexprssionbilayer}
\overline \didv = \frac{G_0}{\nu_0 A}\sum_{\mathbf k}\left[
 \vk^2 \deltazero[-] + 
 \uk^2 \deltazero[+]
\right].
\end{equation}
Here, we also include positive biases, for which the tunneling conductance involves $|\uk|^2$. Figures \ref{fig:figfour}a,b show the spatially averaged conductance for two interlayer distances. As discussed above, we observe a satellite peak emerging at negative bias, which arises from the $|\vk|^2$ term in Eq.\ \eqref{eq:didvstatepexprssionbilayer}. Increasing $|z_\cond-z_\val|$ reduces the threshold $E_G$ necessary for the onset of exciton condensation \cite{macdonaldWuTheoryTwodimensionalSpatially2015}, so that the satellite peak and the conduction band
move closer together, as seen by comparing Figs.~\ref{fig:figfour} a and b.

Integrating $\overline \didv$ across the satellite peak provides a measurement of the local exciton density in the ground state,
\begin{equation}
\label{eq:peakintegralnxconversion}
 \overline I (V= -\infty) = \int^{-\infty}_{E_{-}^0} dV \overline \didv =
- G_0  \frac{\rydberg}{e} 4 \pi \nx .
\end{equation}
This may be particularly useful in inhomogeneous samples, where $\nx$ is spatially varying, so that global measurements have an inherently limited resolution. We note that the  averaging in Eq.\ \eqref{eq:peakintegralnxconversion} serves to eliminate intra-unit-cell structure, so that it is consistent with inhomogeneities on scales large compared to the lattice spacing. 

\section{Spin- and valley degrees of freedom}
\label{sec:spinful}

\subsection{Minimal model of a spinful exciton insulator}

A minimal model for spinful exciton insulators \cite{riceHalperinExcitonicStateSemiconductorSemimetal1968} has spin degenerate conduction and valence bands with independent spin-rotation symmetries. This can be thought of as two copies of the model of Sec.~\ref{sec:hf}. Time reversal symmetry forces $\mathbf w_\cond$ and $\mathbf w_\val$ to be time-reversal invariant momenta, so that the momentum difference $\mathbf w = \mathbf w_\cond - \mathbf w_\val$ is also a time-reversal invariant momentum, which we assume to be nonzero.  Using the mean-field ground state for the spinless case, Eq.\ \eqref{eq:excitongs}, one mean-field charge neutral ground state for this spinful model is simply two copies of the spinless solution, 
\begin{equation}
\label{eq:excitongsspinful}
\GS =\Pi_{\mathbf{k},s=\uparrow,\downarrow}(
        \uk + v_{\mathbf{k}}  c_{\mathbf{k},\cond,s}^{\dagger} c_{\mathbf{k},\val,s}    
        )\ehvacuum ,
        \end{equation}
which has doubly degenerate upper and lower bands with energies given by Eqs.\ \eqref{eq:bandenergiesdef}.

Using the basis
$\left\{ \ket{\mathbf k,\cond, \uparrow},
\ket{\mathbf k,\cond, \downarrow}
\ket{\mathbf k,\val, \downarrow}
\ket{\mathbf k,\val, \downarrow}
\right\}$, the projector onto the
spin-degenerate upper and lower bands is given by a tensor product, 
\begin{eqnarray}
\label{eq:projectorupperspin}
    P_{\mathbb{1}}^{\pm}(\mathbf k) &=&
    P^\pm(\mathbf k) \otimes \mathbb{1}.
    \end{eqnarray}
Explicitly, we have for the upper-band projector
\begin{eqnarray}
P_{\mathbb{1}}^{+}(\mathbf k) &=&
    \begin{pmatrix}
        (\uk)^2 & 0&-\uk \vk&0 \\
        0&(\uk)^2 & 0&- \uk \vk \\
        - \uk \vk &0& (\vk)^2&0\\
        0&- \uk \vk & 0&(\vk) ^2
    \end{pmatrix},
\end{eqnarray}
and analogously for the lower-band projector. The model has a $U(2)\times U(2)$ symmetry, so that other ground states can be generated by rotating the spin axes of conduction and valence bands. The ground-state manifold can be generated by the transformation
\begin{equation}
    \begin{pmatrix}
        w & 0 \\
        0& \mathbb{1}
    \end{pmatrix},
\end{equation}
where $w \in U(2)$ rotates the spin axis in the conduction band. The band projectors are now labeled by the matrix $w$ and read
\begin{equation}
\label{eq:projectorwdef}
    P_{w}^{\pm}(\mathbf k) =
    \begin{pmatrix}
        w & 0 \\
        0& \mathbb{1}
    \end{pmatrix}
    P_{\mathbb{1}}^{\pm}(\mathbf k)
    \begin{pmatrix}
        w^\dagger & 0 \\
        0& \mathbb{1}
    \end{pmatrix}.
\end{equation}

By analyzing the energetics for different matrices in the presence of terms beyond the dominant term approximation,
Halperin and Rice \cite{riceHalperinExcitonicStateSemiconductorSemimetal1968} 
find two candidate ground states, spin-density wave (SDW) and charge-density wave (CDW), described by
\begin{eqnarray}
w_{\text{SDW}}&=& 
\mathbf n \cdot \boldsymbol{\sigma} \\ 
w_{\text{CDW}}&=& 
\mathbb{1}.
\end{eqnarray}
The vector $\mathbf n$ gives the spin direction of the SDW. For weak (strong) electron-phonon coupling, the SDW (CDW) state is the ground state \cite{riceHalperinExcitonicStateSemiconductorSemimetal1968}.

\subsection{Spin-resolved tunneling}

The spin structure can be probed by a tip that is partially spin polarized along a certain direction, with spin-resolved densities of states 
\begin{eqnarray}
\label{eq:dosspinpol}
\nu_\uparrow = \nu_t + \delta \nu / 2 \\
\nu_\downarrow  = \nu_t - \delta \nu /2.
\end{eqnarray}
We introduce electron operators  
\begin{equation}
\psi_s^\dagger(\rtip) = \sum_{\mathbf k}\sum_{i}\phi_{\mathbf k,i}(\rtip,s) c^\dagger_{\mathbf k,i}  
\end{equation}
for the two (pseudo-)spin species $s=\uparrow/\downarrow$ of tip electrons and let the band label $i$ also run over the spin labels. As the tunneling currents of the two spin species add, we define separate Bloch-function factors
$[\wfvector_{s}(\mathbf r)]_{i,j}=\auc e^{i(\mathbf w_j-\mathbf w_i)\cdot \mathbf r}  u^*_{\mathbf w_i,i}(\mathbf r,s)u_{\mathbf w_j,j}(\mathbf r,s)$ for  each spin species.

We now separate the tunneling current into its contributions proportional to the average density of states $\nu_t$ as well as the density-of-states difference,
\begin{multline}
\label{eq:didvstatepexprssionspin}
\didv = \frac{G_0}{\nu_0}
\Tr \Biggl\{ \frac{1}{A}\sum_{\mathbf k, \indexsolution} P^{(\indexsolution)}(\mathbf k)
\deltazero
 \\
\times \left[ (\wfvector_\uparrow(\mathbf r)+\wfvector_\downarrow(\mathbf r))+
 \frac{\delta\nu}{2\nu_t}(\wfvector_\uparrow(\mathbf r)-\wfvector_\downarrow(\mathbf r))\right]  \Biggr\}.
\end{multline}
The term proportional to $\delta \nu$ probes the spin structure of the exciton condensate. In this section, we define $G_0$ without the spin-degeneracy factor $g_s$.

\subsection{Tunneling into the spinful model}

For a tip spin polarized along $\mathbf m$, the matrix of wavefunctions is [see Eq.\ \eqref{eq:wfvectorspinless}]
\begin{equation}
\wfvector_\uparrow (\mathbf r) +
\wfvector_\downarrow (\mathbf r)= 
\wfvector^{\text{spinless}} (\mathbf r)  \otimes \mathbb{1},
\end{equation}
for spin-averaged tunneling, and
\begin{equation}
\wfvector_\uparrow (\mathbf r) -
\wfvector_\downarrow (\mathbf r)= 
\wfvector^{\text{spinless}} (\mathbf r)  \otimes \mathbf m \cdot \boldsymbol{\sigma}.
\end{equation}
for the spin-polarized tunneling contribution. Here, we use the notation $\Phi^\mathrm{spinless}$ for the Bloch-function factor in the spinless case, see Eq.\ \eqref{eq:spinlessPhi}. With this, the traces in Eq.~\eqref{eq:didvstatepexprssionspin} can be readily evaluated. 

First, consider the spin-averaged tunneling. The spatial average recovers the spinless answer, measuring the total density of states as discussed in Sec.\ \ref{sec:monolayer}. In addition, there will be spatial oscillations with wavevector $\mathbf w$. However, due to time-reversal symmetry, $- \mathbf w$ and $\mathbf w$ are equivalent points in the Brillouin zone. Thus, both offdiagonal components in Eq.\ \eqref{eq:wfvectorspinless} contribute. Accounting for this, the result for the spatially oscillating, spin-averaged component is
\begin{multline}
\label{eq:didvstatepexprssionmonolayerwspinful}
\left.\didv\right|_{\mathbf w} =G_0 \frac{1}{\nu_0} \braket{{u_{\val}|u_{\cond}}} 
\Tr\left(w + w^\dagger\right)
\\ \times
\frac{1}{A}\sum_{\mathbf k}
\uk \vk 
 \left[\deltazero[-] - \deltazero[+]\right],
\end{multline}
Compared to the spinless model absent time-reversal symmetry, we have the extra factor $\Tr\left(w + w^\dagger\right)$. For the SDW state, this factor vanishes as the oscillations of the up and down spin exactly cancel. For the CDW state, on the other hand, we recover two times the answer of Sec.\ \ref{subsec:monolayerosc}. The additional factor of two can be traced back to $-\mathbf w $ and $\mathbf w$ being equivalent momenta in this time-reversal invariant model. As in Sec.\ \ref{subsec:monolayerrecovery}, the exciton  wavefunction can be recovered.

The spin-polarized component can be evaluated ana\-logously. The spatial average vanishes for all the states considered, as none of them has an overall magnetization, differing only in their interband spin correlations. There is again a spatially oscillatory component with wavevector $\mathbf{w}$, which reads
\begin{multline}
\label{eq:didvstatepexprssionmonolayerwspinfull}
\left.\didv\right|_{\mathbf w } = G_0\frac{1}{\nu_0}\frac{\delta \nu}{2\nu_t}\braket{u_\val|u_{\cond}} 
\Tr\left[\mathbf m \cdot \boldsymbol{\sigma} (w + w^\dagger)\right]
\\
\times \frac{1}{A}\sum_{\mathbf k}
\uk \vk 
 \left[\deltazero[-] - \deltazero[+]\right].
\end{multline}
This is nonvanishing only for the SDW order parameter, for which we have $\Tr\left[\mathbf m \cdot \boldsymbol{\sigma} (w + w^\dagger)\right] = 2 \mathbf m \cdot \mathbf n $. The oscillating signal is maximal when the tip magnetization is aligned with the SDW direction. Again, these oscillations allow the recovery of the exciton wavefunction, according to the recipe of Sec.\ \ref{subsec:monolayerrecovery}.

\subsection{Extension to multivalley systems}

Systems with multiple valleys can be treated along the same lines as single valley systems, albeit with an enlarged space of spin-valley orderings. At the level of the dominant-term approximation, the ground state can be described as a pairing between conduction and valence band flavors. If the number of conduction and valence-band valleys does not match, 
some flavors have to remain unpaired, and as a result, exciton condensates are only stable in the BEC regime even at the mean-field level \cite{riceHalperinExcitonicStateSemiconductorSemimetal1968}.

Generically, pairing between conduction and valence bands offset by a finite momentum will lead to spatial oscillations. As an example, consider the recently studied exciton insulator candidate \wte{} \cite{wuJiaEvidenceMonolayerExcitonic2022a,cobdenSunEvidenceEquilibriumExciton2022} which has two conduction and a single valence band valley per spin species. An analysis of the possible orders \cite{parameswaranKwanTheoryCompetingExcitonic2021,mooreWangBreakdownHelicalEdge2023} suggests two candidate ground state orders -- a spin spiral and a SDW. Both are predicted to possess spatially dependent spin ordering, which should be observable in spin-resolved STM and allow for the recovery of wavefunctions as detailed in Sec.\ \ref{subsec:monolayerrecovery}. More generally, however, it is in principle possible that the spatial oscillations cancel out in the total as well as the spin-resolved tunneling conductance due to the multiple paired flavors. 

\section{Discussion}
\label{sec:discussion}

Tunneling experiments have long been central to the study of superconductivity. In view of the close analogies between superconductors and excitonic insulators, it is natural that tunneling experiments will also play an important role in exploring the latter. Tunneling into excitonic insulators in the limit of high exciton density (BCS limit) is closely analogous to superconductors. The formation of an excitonic insulator is signaled by the opening of a gap enclosed by divergent coherence peaks. However, this gap is expected to be strongly suppressed by screening. 

Here, we focused on the limit of low exciton densities (BEC limit), a regime which is less explored also for superconductors. In this regime, exciton formation is no longer signaled by a gap opening as even in the absence of exciton condensation, the system is a narrow-gap semiconductor. Exciton condensation merely leads to a small increase of this gap, and the emergence of nondiverging peaks at the edges of the excitation gap. Nevertheless, we found that there are tunneling characteristics of the formation of an exciton insulator. 

In monolayers, exciton condensation is signaled by a spatially oscillating contribution to the tunneling conductance, which breaks the underlying lattice symmetry. We showed how in the limit of low exciton density, measurements of this spatially oscillating contribution along with the spatially averaged tunneling conductance can be used to extract the exciton wavefunction. While spatial oscillations would also exist in the BCS limit, extracting the exciton wavefunction is limited to the BEC limit. 

In bilayer samples, exciton formation is signaled by a characteristic peak. This peak directly reflects the presence of excitons, but is insensitive to the coherence of the condensate. Its integrated strength can be used to extract the density of excitons in the sample. 

These results emphasize that STM measurements are well suited to probe exciton condensation. They enable the extraction of important information like the exciton density and exciton wavefunctions. While exciton density can also be measured by alternate means, STM measurements allow it to be measured locally. STM measurements should also be valuable to probe localized in-gap states in excitonic insulators
\cite{kwon2025yushibarusinovboundstatesexciton}. Moreover, they can be used to exploit sample inhomogeneity, for instance by focusing on low-disorder regions suitable for exciton condensation, a major advantage compared to spatially averaging techniques such as angle-resolved photo-emission spectroscopy \cite{forroCercellierEvidenceExcitonicInsulator2007,ohtaSekiExcitonicBoseEinsteinCondensation2014,chenGaoEvidenceHightemperatureExciton2023}. This may become especially powerful when combined with intentional sample engineering, enabling a wide range of model parameters to be studied in a single device \cite{wangZhangEngineeringCorrelatedInsulators2024,Gu2024Remote}. 

An interesting open question is the effect of finite temperature
\cite{novoselovFoglerHightemperatureSuperfluidityIndirect2014}.
At finite temperature, fermionic mean-field in the BEC regime breaks down \cite{nozieresComteExcitonBoseCondensation1982,littlewoodZhuElectronholeSystemRevisited1996,yoshiokaAsanoExcitonMottPhysics2014}, failing to account for the center-of-mass motion of excitons. Upon increasing temperature, the exciton condensate is expected to undergo a Berezinskii-Kosterlitz-Thouless transition, and eventually turn into a classical exciton gas \cite{novoselovFoglerHightemperatureSuperfluidityIndirect2014}. In a monolayer, the lattice-symmetry breaking oscillatory components of $\didv$ will disappear at the transition, as quasi-long range order is lost beyond the  transition. In contrast, in a bilayer, the satellite peak should stay, even if thermally broadened, so long as excitons are present. 
The numerical data used to generate the figures of this article are available from Zenodo \cite{zenodo}.

\begin{acknowledgments}
We are grateful to Jiewen Xiao, Daniel Mu\~noz-Segovia, Alexander Holleitner, and Leonid Glazman for helpful discussions. Work at Freie Universit\"{a}t Berlin was supported by Deutsche Forschungsgemeinschaft through CRC 183 (project C02) as well as CRC 1772 (project B06). Work at Aalto University was supported by a grant from the Simons Foundation (SFI-MPS-NFS-00006741-12, P.T.) in the Simons Collaboration on New Frontiers in Superconductivity.
\end{acknowledgments}

\appendix

\section{Perturbative expansion in the dilute limit}
\label{app:solutionofselfconsistencyeq}

We solve the self-consistency equation \eqref{eq:selfconsistencyeq} to 
higher orders for small $\vk$. The appropriate dimensionless expansion parameter is $\lambda \bohrradius$, with the exciton Bohr radius 
\begin{equation}
\bohrradius = \frac{4\pi\epsilon_0\epsilon \hbar^2}{e^2m}.
\end{equation}
Physically, $\lambda {\bohrradius} \ll 1$ is equivalent to $\nx (\bohrradius)^2 \ll 1$, the limit of low exciton density.

For simplicity of notation, we write our expansion in powers of $\lambda$. According to Eq.~\eqref{eq:momentumspaceexcitonschroedinger}, $\vk = \lambda \phi_{\mathbf k}$ and $E_G=E_b$ to lowest order. We expand to higher orders as
\begin{align}
    \vk &= \lambda \phi_{\mathbf k} + \lambda^2  \vk^{(2)} + \lambda^3  \vk^{(3)} +\ldots \\
    \uk &= \sqrt{1-\vk ^2}
    \\
    E_G &= E_b  +\lambda  E_G^{(1)}+ \lambda^{2}E_G^{(2)} +\ldots
\end{align}
We choose the functions $\phi_{\mathbf k}$, $\vk^{(2)}$,$\vk^{(3)},\ldots$ to be orthogonal to each other.

To order $\lambda^2$, Eq.~\eqref{eq:selfconsistencyeq} reads
\begin{equation}
\label{eq:selfconsistencyeq2}
E_G^{(1)} \phi_{\mathbf k} + \frac{\hbar^2 \mathbf k ^2}{2m}\vk^{(2)}  -\frac{1}{A}\sum_{\mathbf k'}V^{cv}_{\mathbf k-\mathbf k'}v^{(2)}_{\mathbf k'}=-E_b\vk^{(2)}.
\end{equation}
Projecting onto $\phi_{\mathbf k}$ and using Eq.~\eqref{eq:momentumspaceexcitonschroedinger}, we obtain $E_G^{(1)}=0$. This in turn implies with Eq.~\eqref{eq:momentumspaceexcitonschroedinger} that $\vk^{(2)} \propto \phi_{\mathbf k}$. Hence $\vk^{(2)}$ can be absorbed into $\phi_{\mathbf k}$, so that we set $\vk^{(2)}=0$.

To order $\lambda^3$, Eq.~\eqref{eq:selfconsistencyeq} reads
\begin{align}
\label{eq:selfconsistencyeq3}
&\left[E_G^{(2)} +\frac{1}{A}\sum_{\mathbf k'}\frac{\phi_{\mathbf k'}^2}{C_G}-\frac{2}{A}\sum_{\mathbf k'}(V^{cc}_{\mathbf k-\mathbf k'}\phi_{\mathbf k'}^2 -\phi_{\mathbf k}V^{cc}_{\mathbf k-\mathbf k'}\phi_{\mathbf k'})\right]\phi_{\mathbf k} 
\nonumber \\
& +\left[\left(\frac{\hbar^2 \mathbf k ^2}{2m}+E_b\right)\phi_{\mathbf k}
-\frac{1}{A}\sum_{\mathbf k'}V^{cv}_{\mathbf k-\mathbf k'}\phi_{\mathbf k'}\right]\frac{\phi^2_{\mathbf k}}{2} 
\nonumber \\
& + \frac{\hbar^2 \mathbf k ^2}{2m}\vk^{(3)}  -\frac{1}{A}\sum_{\mathbf k'}V^{cv}_{\mathbf k-\mathbf k'}v^{(3)}_{\mathbf k'}=-E_b\vk^{(3)}.
\end{align}
The second line vanishes by virtue of Eq.~\eqref{eq:momentumspaceexcitonschroedinger}. Multiplying by $\phi_{\mathbf k}$ and summing over $\mathbf k$ removes the $\vk^{(3)}$ term, allowing $E^{(2)}_G$ to be evaluated as
\begin{multline}
\label{eq:gapsecondorder}
E_G^{(2)} = -\frac{1}{C_G}+\frac{2}{A^2}\sum_{\mathbf k,\mathbf k'}(V^{cc}_{\mathbf k-\mathbf k'}\phi_{\mathbf k}^2\phi_{\mathbf k'}^2 -\phi^3_{\mathbf k}V^{cv}_{\mathbf k-\mathbf k'}\phi_{\mathbf k'}).
\end{multline}
Here, we used the normalization of  $\phi_{\mathbf k}$. Noting that $\nx = \lambda^2 + O(\lambda^6)$, the exciton compressibility
$\frac{d\mux}{d\nx} = -\frac{dE_G}{d\nx} = -E_G^{(2)}$ is given by
\begin{multline}
\label{eq:dmudxexpression}
\frac{d\mux}{d\nx}= \frac{1}{C_G}-\frac{2}{A^2}\sum_{\mathbf k,\mathbf k'}(V^{cc}_{\mathbf k-\mathbf k'}\phi_{\mathbf k}^2\phi_{\mathbf k'}^2 -\phi^3_{\mathbf k}V^{cv}_{\mathbf k-\mathbf k'}\phi_{\mathbf k'}).
\end{multline}
The first term is due to the geometric capacitance and reflects the dipole-dipole repulsion of excitons in bilayers. 
The second term is the gain of intraband exchange energy, while the last term is the inter-band exchange repulsion. For a monolayer, we can use the analytical form of the wavefunction, Eq.~\eqref{eq:excitonwfmonolayer} to perform the sums over $\mathbf k$ and $ \mathbf k'$. We obtain $\frac{d\mux}{d\nx}\approx 6.0566\, \rydberg (\bohrradius)^2$.

\section{Numerical details}
\label{app:numerical}

Our numerical results for the coefficients $\uk, \vk$ are obtained by solving the mean-field Hamiltonian Eq.\ \eqref{eq:HBdG} self-consistently. The solutions satisfy Eq.\ \eqref{eq:selfconsistencyeq}. 
We use a finite momentum grid in the radial direction, assuming that the solution is spherically symmetric ($s$-wave). For the monolayer case, special care needs to be taken to ensure proper convergence in view of the singular nature of the Coulomb potential. We choose grids that are sufficiently dense and large so that the exciton density is within $2\%$ of the prediction of Eq.~\eqref{eq:dmudxexpression}, i.e., $\mux = 6.0566 \rydberg (\bohrradius)^2 \nx$. Specifically, we chose 
a grid of $N_{k} = 6839$ points going up to $k_{max}=180 \frac{1}{\bohrradius}$ for Figs.\ \ref{fig:figtwo} and \ref{fig:figthree}). For the bilayer calculation presented in Fig.\ \ref{fig:figfour}, we use $N_{k} = 1139$ points going up to $k_{max}=60 \frac{1}{\bohrradius}$. The exact grids are included in the dataset on Zenodo \cite{zenodo}.


%

\end{document}